\documentclass{aa}  

\usepackage{graphicx}
\usepackage{txfonts}
\usepackage[breaklinks=true]{hyperref}
\usepackage{natbib} 
\bibpunct{(}{)}{;}{a}{}{,} 

\usepackage{verbatim}
\usepackage{url}
\usepackage{epstopdf}

\begin{document}

   \title{A statistical test on the reliability of the non-coevality of stars in binary systems\thanks{On-line calculator available at http://astro.df.unipi.it/stellar-models/W/} 
     }

   \subtitle{}

   \author{G. Valle \inst{1,2,3}, M. Dell'Omodarme \inst{3}, P.G. Prada Moroni
     \inst{2,3}, S. Degl'Innocenti \inst{2,3} 
          }

   \authorrunning{Valle, G. et al.}

   \institute{
INAF - Osservatorio Astronomico di Collurania, Via Maggini, I-64100, Teramo, Italy 
\and
 INFN,
 Sezione di Pisa, Largo Pontecorvo 3, I-56127, Pisa, Italy
\and
Dipartimento di Fisica ``Enrico Fermi'',
Universit\`a di Pisa, Largo Pontecorvo 3, I-56127, Pisa, Italy
 }

   \offprints{G. Valle, valle@df.unipi.it}

   \date{Received 25/03/2015; accepted 30/12/2015}

  \abstract
   {}
   {  
We develop a statistical test on the expected difference in age estimates of two coeval stars in detached double-lined eclipsing binary systems that are only caused by observational uncertainties. 
We focus on stars in the mass range [0.8; 1.6] $M_{\sun}$, with an initial metallicity [Fe/H] from -0.55 to 0.55 dex, and on
stars in the main-sequence phase.
}
{       
The ages were obtained by means of the SCEPtER technique, a maximum-likelihood procedure relying on a pre-computed grid of stellar models. The observational 
constraints used in the recovery procedure are stellar mass, radius, effective temperature, and metallicity [Fe/H]. To check the effect of the 
uncertainties affecting observations on the (non-)coevality assessment, the chosen observational constraints were subjected to a Gaussian perturbation before applying the 
 SCEPtER code. We defined the statistic $W$ computed as the ratio of the absolute difference of estimated ages for the two stars over the age of the older one. We determined
  the critical values of this statistics above which coevality can be rejected in dependence on the mass of the two stars, on the initial metallicity [Fe/H], and on the evolutionary stage of the primary star.
}
  {
The median expected difference in the reconstructed age between the coeval stars of a binary system --   caused alone by the observational uncertainties -- shows a strong dependence on the evolutionary stage. 
This ranges from about 20\% for an evolved primary star to about 75\% for a near ZAMS primary. The median difference also shows an increase with the mass of the primary star from 20\% for 0.8 $M_{\sun}$ stars to about 50\% for 1.6 $M_{\sun}$ stars. 
The reliability of these results was checked by repeating the process with a grid of stellar models computed by a different evolutionary code; the median difference in the critical values was only 0.01.
We show that the $W$ test is much more sensible to age differences in the binary system components than the alternative approach of comparing the confidence interval of the age of the two stars.
We also found that the distribution of $W$ is, for almost all the examined cases, well approximated by beta distributions. 
 }
{
The proposed method improves upon the techniques that are commonly adopted for judging the coevality of an observed system. It also provides a result founded on reliable
statistics that simultaneously accounts for all the observational uncertainties.
}

   \keywords{
Binaries: eclipsing --
methods: statistical --
stars: evolution --
stars: low-mass 
}

   \maketitle

\section{Introduction}\label{sec:intro}

The impossibility to simultaneously fit 
the observed characteristics of stars in a double-lined detached binary 
system  with a single isochrone is usually interpreted as the need to introduce some modifications in
the current generation of evolutionary codes, such as 
adjustments of the external convection efficiency through the mixing-length parameter
\citep[e.g.][]{Torres2006,Clausen2009,Morales2009}. 
Moreover,  these
systems are often adopted in studies on the calibration of convective
core
overshooting by fine-tuning the isochrone fit \citep[see, among many,][]{Andersen1990,Ribas2000,Lacy2008,Clausen2010}.  

However, an apparent age difference between the binary system components could
simply arise by fluctuations that arise as a result of the uncertainties in the observational
constraints adopted in the estimation. This was already pointed out in the literature;
as an example, \citet{Gennaro2012} conducted some tests for nine combinations of stellar masses on pre-main-sequence stars. More recently, \citet{binary} presented some basic results about the expected
differences in the age estimates of the two coeval members in double-lined detached binary
systems, obtained by the grid-based technique SCEPtER \citep{scepter1, eta}. We found that 
the differences are large and reach a median value of 60\% for a mass ratio of 0.5.

Therefore, given the relevance of the coevality hypothesis, it is of paramount importance to set out accurate methods for evaluating  whenever a detected difference in ages of the two stars in an observed system is 
"too high" to be coeval on statistical grounds. Unfortunately, the recent literature is quite inhomogeneous with regard to this problem. It is common practice to compare the observed system to sets of isochrones to determine the system age, while the best-fit ages for a single star are seldom provided. In most cases the details of the fitting techniques are scarcely discussed. Moreover, it is common to individually obtain the system age estimates  for a set of different metallicities, compatible with the error on the observed [Fe/H], and then to assess the error caused by the metallicity uncertainty by considering the obtained age spread \citep[see among many][]{Clausen2008, Lacy2008, Torres2009, Clausen2010, Rozyczka2011,Sowell2012}. The drawback of this method is that it cannot account for possible error correlations among masses, metallicities, and effective temperatures of the stars.
For age estimation purposes, stellar tracks for the precise observed masses were calculated in other cases \citep[e.g.][]{Vos2012}.
The literature also reports a two-stage fitting method,  which first estimates the best-fit stellar model metallicity that is compatible with the spectroscopic determination, and then finds the best-fitting isochrone for this fixed metallicity \citep[see e.g.][]{ Lacy2010, Welsh2012}. 
Overall, each author relies on different methods and different stellar tracks, computed with chemical and physical input that results in age estimates that sometimes are vastly different. The worst problem is that a reliable and consistent treatment of the errors, at least those arising from the uncertainties in the observed quantities adopted in the fitting, is generally lacking, and a reliable error on the age estimates is rarely
provided.  This makes evaluating the reliability of the deviation from coevality and comparing the results of different authors
very challenging.

The aim of this paper is to partially alleviate this problem
by  providing a statistical test of the coevality hypothesis  that supplements the currently adopted techniques. Our intent is to develop a reliable method that consistently treats the observational errors, and to show its usage in practice.  To this aim, we evaluate
the dependence of the expected age differences on the mass of the two stars,
on their metallicity, and on their relative age\footnote{The relative age is defined as the ratio between the age of the star and the age of 
the same star at central hydrogen depletion (the age is conventionally set to
0 at the ZAMS position).} on the main sequence. 
We provide the critical values of the expected differences in age to be used to asses 
if the reconstructed non-coevality is simply the result of a
random fluctuation.
We also show the differences between the coevality test and the commonly adopted comparison of the age confidence interval of the observed stars. 

The work developed in this paper is framed in the theory of the frequentistic statistical hypothesis testing \citep[see among many][for details]{snedecor1989,Feigelson2012}. In the present-day formulation, this sound mathematical theory  is rooted in the first decades of the nineteenth century and is mainly based on the theoretical studies of Pearson, Gosset (Student), Neyman, and Fisher \citep[see e.g.][]{Fisher1925, Neyman1933}. Without any claim at completeness, we recall that the theory requires formulating a scientific hypothesis to check (in our case the stellar coevality) that is often referred to as the null hypothesis $H_0$, to define a parent population on which the hypothesis should be verified, to define a statistic $\cal{T}$ to be adopted in the hypothesis testing, and to derive the distribution of $\cal{T}$ under $H_0$. From this distribution is then identified a rejection region of $H_0$; for an unilateral test (as the one in the present paper) this traditionally corresponds to values of the statistics above the 95th quantile of the distribution (the value corresponding to the 95th quantile is called critical value). By this choice the experimenter sets the so-called level $\alpha$ of the test to a value of 0.05, corresponding to the probability of type I errors (i.e. rejecting $H_0$ when it is indeed true).     
After these theoretical derivations, the test can be applied to a random sample drawn from the parent population to verify $H_0$. The reference statistic $\cal{T}$ is computed on this sample and compared with the chosen quantile of its distribution. A sample value of $\cal{T}$ above the critical value leads to rejecting $H_0$.

\section{Methods}\label{sec:method}

Stellar ages were determined by means of the SCEPtER pipeline, a maximum-likelihood technique 
relying on a pre-computed grid of stellar models and a set of observational constraints \citep[see e.g.][]{eta}. 
A first application to eclipsing binary systems has been extensively described in
\citet{binary}. The grid of models covers the evolution from the zero-age main-sequence (ZAMS) 
up to the central hydrogen depletion of stars with masses in the range [0.8; 1.6] $M_{\sun}$ and initial metallicities [Fe/H] from $-0.55$ dex to 0.55 dex. The grid was computed by means of the FRANEC stellar evolutionary code \citep{scilla2008, tognelli2011}, in the same
configuration adopted to compute the Pisa Stellar
Evolution Data Base\footnote{\url{http://astro.df.unipi.it/stellar-models/}} 
for low-mass stars \citep{database2012, stellar}. 
The details of the standard grid of stellar models, the sampling procedure, and the age estimation technique are fully described in 
\citet{eta} and \citet{binary}, while the adopted input and the related uncertainties are discussed in  \citet{cefeidi, incertezze1, incertezze2}.

To test the ability of grid-based maximum-likelihood techniques of assessing 
the coevality of binary components within random error fluctuation, we chose the most favourable 
scenario, that is, the case in which stellar models perfectly agree 
with observations and the binary members are coeval by construction. 
To achieve this optimistic situation, we built a synthetic dataset by sampling the artificial 
binary systems from the same grid of stellar models as was used in the recovery procedure. 
The effect on the coevality assessment of the current typical observational errors was 
simulated by adding random Gaussian perturbations -- with $ \sigma $ equal to the uncertainty --
  to the chosen observational constraints. 

As already mentioned, the binary components are coeval by construction
because they have been generated with the same age 
in the sampling stage. However, as a consequence of the perturbation procedure mimicking the observational errors, 
the recovery might provide two different age estimates for the binary members. 

Our aim is to devise a criterion founded on statistics to evaluate the likelihood that the recovered 
non-coevality is genuine and not merely the result of observational errors. 
More formally, the null hypothesis $H_0$ for which we wish to develop a statistical test is that the stars are coeval within the random variability caused by the observational errors.
We aim to define a statistics to test $H_0$, obtaining a
rejection region at level $\alpha,$ and 
thus a critical value identifying it. We define $A_1$ and $A_2$ as the estimated ages of the 
two members, with $A_1 > A_2$. We focus on the statistics $W$ defined as 
\begin{equation}
W = \frac{A_1-A_2}{A_1}
.\end{equation}
$W$  then varies from 0 (when the stars are estimated to be coeval) to
 1 (when $A_1 \gg A_2$). 
 High values of $W$ lead to the rejection of the null hypothesis, implying that it is very unlikely that standard 
 stellar models might account for the coevality of the stars with the assumed
input or parameters. The point is to establish the critical values above which the coevality rejection is 
statistically significant.

\begin{figure*}
        \centering
        \includegraphics[width=17.5cm,angle=0]{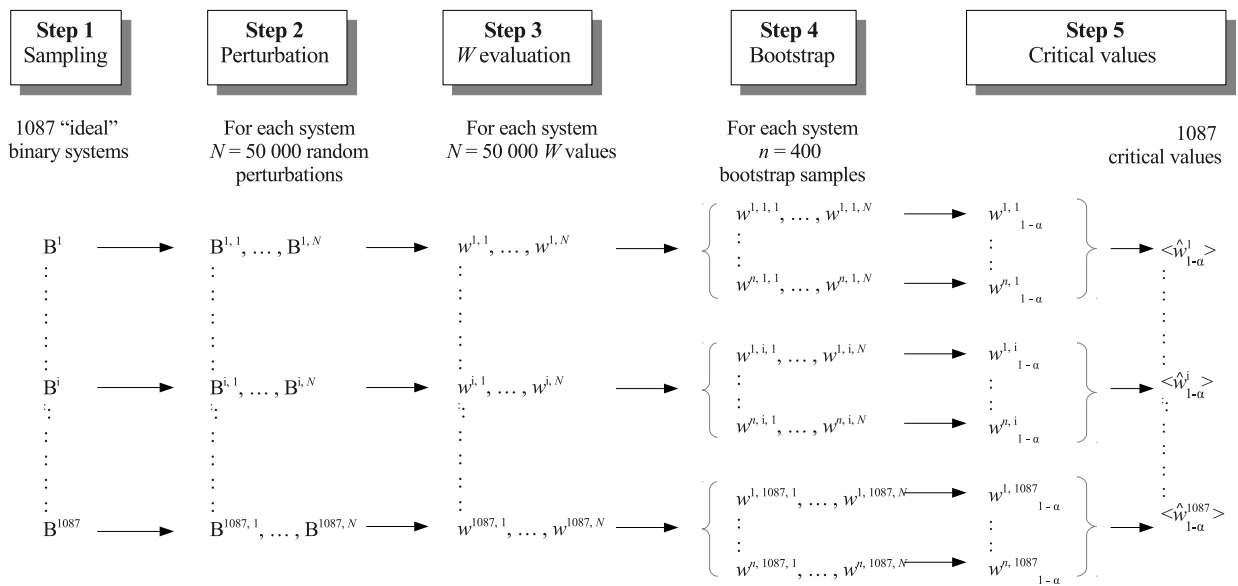}
        \caption{Outline of the design to compute the critical values. Steps 1-3 are described in Sect. ~\ref{sec:sampling}, steps 4-5 in Sect.~\ref{sec:critical}.}
        \label{fig:schema1}
\end{figure*}

To do this, we evaluated the distribution of $W$ -- under the coevality hypothesis $H_0$  -- that arises from observational uncertainties alone and computed the {\it \textup{critical values}}  that
define the range of values $W$ that are compatible with the null hypothesis itself.   
As stated in the introduction, the 95th  quantile of the distribution of the statistics under examination are typically adopted as critical values.
The choice of the quantile defines the level $\alpha$ of the test. 
In the following we assume a level of the test $\alpha = 0.05$ and compute the critical values  $W_{1-\alpha}$. 
For $W > W_{1-\alpha}$ the null hypothesis $H_0$, that is, the coevality of the binary components, is rejected.

\subsection{Sampling and critical values estimation}\label{sec:sampling}

Since the critical value $W_{1-\alpha}$ depends on the characteristics of the binary system, it is not 
possible to provide a single value suitable for all the possible binary configurations. 
To describe the dependence of $W_{1-\alpha}$ on the physical 
parameters that identify the binary system, a long and time consuming procedure is required. 
Figure ~\ref{fig:schema1} outlines the steps -- described in detail below -- that we followed in the estimation procedure. 

\paragraph{Step 1}

To produce a representative sample of synthetic binary systems spanning the possible combinations of masses, metallicity, and relative ages of the 
two stars, we performed a systematic sampling from the standard grid of stellar models. 
We selected a set of nine stellar masses from 0.8 $M_{\sun}$ to 1.6 $M_{\sun}$
with a step of 0.1 $M_{\sun}$. The metallicity was restricted to initial [Fe/H]
values of $-0.55$, $-0.25$, 0.0, 0.25 and 0.55. The relative age $r$ of the primary
star was chosen to assume values from 0.1 to 0.9 with a step of 0.2.
Then, for each value of the primary mass $M_1$, metallicity [Fe/H], and
relative age, 
we selected the corresponding model from the grid. Linear interpolation  in
effective temperature, radius, and metallicity was performed to obtain
a stellar model of the exact chosen relative age. This step fixes the age of the binary
system to $A,$ that is, the age of the selected star.
This synthetic star is coupled to all the
possible secondary components with masses lower than or equal to $M_1$, the same initial [Fe/H], and
age exactly equal to $A$. Linear interpolation in effective
temperature, radius, and metallicity was performed to match
these requirements.
The described sampling scheme produced 1125 possible combinations. However,
since the grid does not contain models in the pre-main-sequence phase, not all these combinations correspond to existing pairs in the grid. 
For example, a 1.6 $M_{\sun}$ model at relative age 0.1 is too young
 to be matched by any model of 0.8 $M_{\sun}$, which is
still in the pre-main-sequence phase. 
 
In conclusion, we found that a total number of 1087
binary systems are possible, and we estimated the corresponding critical values $W_{1-\alpha}$ for each of them.
  
\paragraph{Step 2}

To simulate observational uncertainties, the 1087 synthetic binary systems were subjected to a Gaussian perturbation of all the observed quantities 
with standard deviations of 100 K in $T_{\rm eff}$, 0.1 dex in [Fe/H], 1\% in mass, and 0.5\% in radius. To simulate realistic uncertainties, 
we assumed a correlation of 0.95 between the primary and secondary effective temperatures, 0.95 between the metallicities, 0.8 between the masses, 
and no correlation between the radii. A detailed discussion of these choices and their effect on the final results, also including the motivations for the zero correlation between the radii, can be found in \citet{binary}. 

As a result, for each of these 1087 cases we produced a set of $N$ perturbed systems, where the actual value of $N$ was properly evaluated in steps 4 and 5.

\paragraph{Step 3}

For each of the 1087 analysed systems, we computed the $N$ values of the statistic $W$ corresponding to all the perturbed objects. Finally, we estimated from these 1087
samples the $W_{1-\alpha}(M_1, M_2, {\rm [Fe/H]}, r)$ quantiles, which
approximate the required critical values for the statistical test. Critical values for masses, metallicities, and relative ages different from those adopted in the computations can be obtained by interpolation. 

At the end of this step, we obtained the required critical values. However, the procedure 
cannot stop here. In fact, a different simulation run will produce different values; it is therefore mandatory to estimate the random variability of the results. This is the purpose of the following section.

\subsection{Monte Carlo variability on the critical values}\label{sec:critical}

The accuracy of the computed quantiles $W_{0.95}$ depends on the size $N$ of the Monte Carlo sample, and it is obvious that a larger sample would provide a more accurate estimate. Our aim is to reach a relative precision of 1\% on the estimated critical values; therefore we have to evaluate the minimum $N$ needed to meet this requirement.

More technically, given a sample  $X_1, \ldots, X_N$ of size $N$, taken from a distribution $F$
and density $f$,
we aimed to estimate the $1-\alpha$ quantile, defined as the value for which the distribution function assumes value $1-\alpha,$ that is,
\begin{equation}
W_{1-\alpha} = F^{-1}(1-\alpha).
\end{equation}

\begin{figure}
        \centering
        \includegraphics[height=8cm,angle=-90]{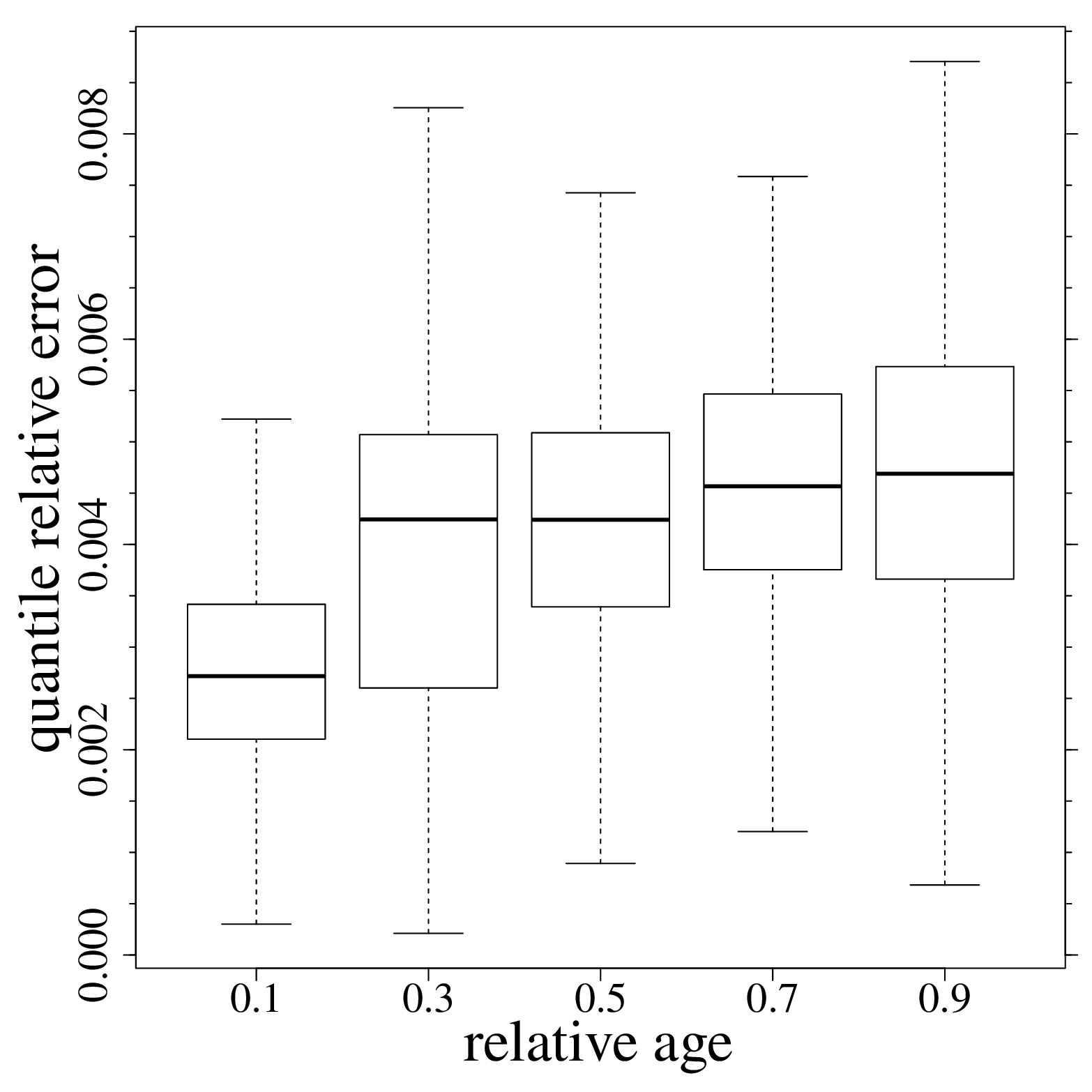}
        \caption{Boxplot of the relative error on the $W_{0.95}$ quantile estimates as a
                function of the primary relative age. Error estimates are obtained by
                bootstrap resampling (see text).}
        \label{fig:error-crit}
\end{figure}

Standard theory establishes that the $1-\alpha$ sample quantile
$X_{\lceil N (1-\alpha) \rceil}$ (the $\lceil N (1-\alpha) \rceil$th smallest observation in
the sample $X_1, \ldots, X_N$) is a consistent estimator\footnote{A
  consistent estimator is an estimator that converges in probability to the
  value to be estimated as the sample size goes to infinity.} of the real quantile. Moreover, the  
asymptotic expansion for the
variance $\sigma^2_N$ of $X_{\lceil N (1-\alpha) \rceil}$ is \citep[see e.g.][]{Stuart1994} 
\begin{equation}   
\sigma^2_N = N^{-1} \alpha (1- \alpha) \; f\left(W_{1-\alpha}\right)^{-2}
+o(N^{-1})
\label{eq:asymvar}  
,\end{equation}
which shows that the convergence rate of Monte Carlo quantile simulations is
$1/\sqrt{N}$.
Unfortunately, Eq.~(\ref{eq:asymvar}) cannot be used for direct computations
since it depends 
on the unknown density function $f(W_{1-\alpha})$.
An usual way to circumvent the problem is to relie on bootstrap estimates
\citep[see e.g][]{Davison1997}. 

\paragraph{Steps 4 and 5}
The approach requires sampling with replacement a number $n$ of samples of size
$N$ from the original $W$ values. For each of these samples, we computed the sample
quantile $\hat w_{1-\alpha} = \{X_{\lceil N (1-\alpha) \rceil}^j$,
$j=1,\ldots,n\}$ and then estimated the arithmetic mean $\left<\hat
w_{1-\alpha}\right> = 1/n \sum_j \hat w_{1-\alpha}^j$ and the variance $\sigma^2_N$ by the 
variance of the bootstrap quantiles, that is,
\begin{equation}
\hat \sigma^2_N = {\rm Var}(\hat w_{1-\alpha}).
\label{eq:bootvar}
\end{equation}
For our computations we adopted $n = 400$.

\citet{Hall1988} proved that $\hat \sigma^2_N/\sigma^2_N = 1 + O(N^{-1/4})$,
which means that the convergence to the true estimator is relatively slow but
sufficient for our purposes. The technique also allows estimating the bias
$b$ of 
the sample $w_{1-\alpha}$
\begin{equation}
b =  X_{\lceil N (1-\alpha) \rceil} - \left<\hat w_{1-\alpha}\right>.
\end{equation}
The sample estimate can therefore be corrected to account for the bias assuming as best quantile estimate the value $\left<\hat
w_{1-\alpha}\right>$.

We therefore performed some exploratory computations to quantify the 
values from Eq.~(\ref{eq:bootvar}) for different $N$. As a result, we
found that $N = 50\,000$ allowed us to reach the required 1\% relative accuracy for all the sample.
The obtained relative errors $\hat \sigma_N/\left<\hat
w_{1-\alpha}\right>$ on the 1087 quantile estimates are shown in
Fig.~\ref{fig:error-crit}, in dependence on the relative age of the binary
system. The figure presents the boxplot\footnote{A boxplot is a convenient
way to  summarize a distribution. The black
thick line marks the median of the distribution, while the box extends from
the 25th to the 75th quantile. The whiskers extend to the extreme values, but
their lengths are limited to 1.5 times the width of the box. Points
outside the extension of the whiskers are omitted from the plot.} of the errors for each relative age. 
 It is apparent
that the median relative errors on the estimated quantile are lower than 0.5\%
for all the explored relative ages.

\section{Critical values $W_{0.95}$}\label{sec:results}

\begin{figure*}
\centering
\includegraphics[height=17cm,angle=-90]{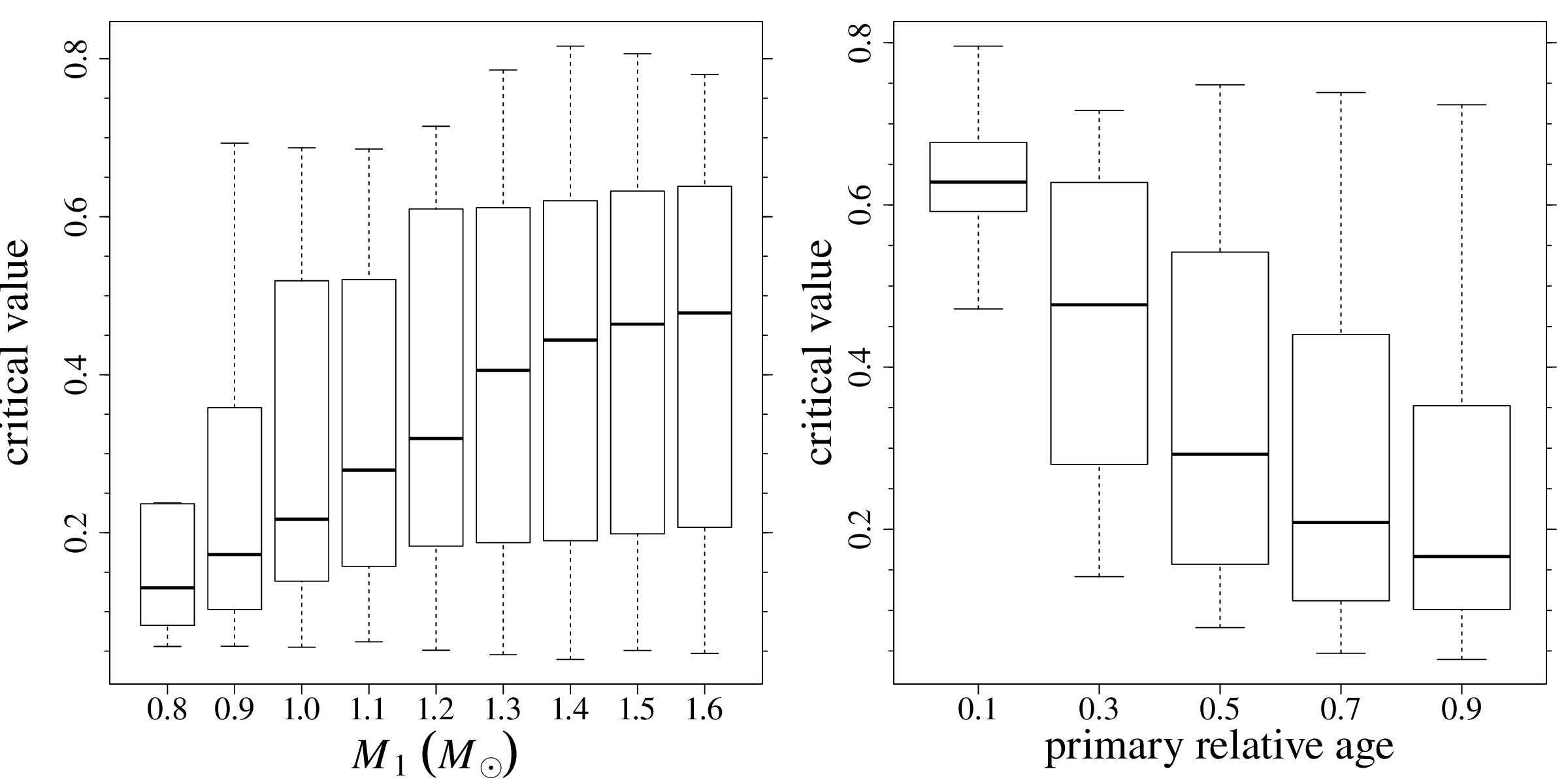}
\caption{{\it Left}: boxplot of the $W_{0.95}$ quantile estimates as a
  function of the mass of the primary star. {\it Right}: same as the {\it \textup{left panel}}, but
  for the dependence on the relative age of the primary star.}
\label{fig:crit-m1-pcage}
\end{figure*}

The computation of the 1087 critical values described in the previous section required a total of $5.4 \times 10^7$ binary system age estimates. 
The results of this huge set of simulations are presented in Appendix~\ref{sec:appcrit},
in Tables from \ref{tab:crit0.1} to \ref{tab:crit0.9}. Each table collects -- for
values of the primary relative age $r$ from 0.1 to 0.9 with a step of 0.2 -- the
critical values $W_{0.95}$ in dependence on the masses of the stars and their
initial metallicity [Fe/H]. 

To use the test in practice, after
estimating the ages of the two stars and computing the value of the statistics $W$, this must be compared 
with the appropriate critical value in the tables. If the computed $W$ is lower than the critical value $W_{0.95}$, the null hypothesis of coevality cannot be rejected.

One complication arises because the values in Tables~\ref{tab:crit0.1} -
\ref{tab:crit0.9} depend on  the observationally unknown primary
relative age $r$. The problem can be solved by estimating $r$, for example 
by means of the same grid technique as adopted for age estimates (more details on this topic are provided in Sect.~\ref{sec:Wint}).  

Our main result here is that the critical values -- and thus the 
critical age relative differences on age estimates of two stars that are coeval by construction -- are  indeed high despite the high precision reached by the observations.
The overall median of the critical values is 0.36, with an interquartile range of [0.16; 0.61]. 
Restricting the test to binary systems of low or intermediate relative age of the primary star ($r \leq 0.5$) results in a median critical value of 0.60 with an interquartile range of [0.45; 0.65]. 
This shows a general behaviour, which is that the closer the primary star is to the ZAMS, the higher is the critical value $W_{0.95}$ and hence the expected 
difference between the ages of the two binary components.
 
Some trends in the critical values presented in Tables~\ref{tab:crit0.1} - \ref{tab:crit0.9} are apparent and easily understandable. The lowest values of $W_{0.95}$ in each
row or column of the tables are usually found for binary systems composed by equal-mass stars. These
are the only combinations for which the relative age of the two stars are the
same, all the others provide a lower relative age for the secondary. Since it is
known \citep[see the extensive discussion in][]{eta} that 
the relative error in age decreases with relative age, it is straightforward to
conclude that the $W_{0.95}$ values on the diagonal of the tables should be the lowest.
From the same argument it follows that an increase of the critical values is expected farther down in the
columns -- that is, increasing the mass of the primary star -- and leftward in the rows  -- that is, decreasing the mass of the secondary star. 
However, closer to the border of the
tables, the trend is less clear since here some edge effects
\citep[see][for a discussion]{scepter1, eta} become dominant.

The discussed trend of the critical values with
the mass of the primary star is shown in the left panel of Fig.~\ref{fig:crit-m1-pcage}. The medians of the boxes increase monotonically
from the 0.8 $M_{\sun}$ models to those with 1.6 $M_{\sun}$. The large spread of
the boxes is due to the variability of the other parameters, that is, the
secondary mass, the metallicity, and the relative age of the primary star.  
The strong effect of the primary relative age $r$ on critical
values $W_{0.95}$ is shown in the right panel of Fig.~\ref{fig:crit-m1-pcage}. For
low values of $r$ both stars are at a young relative age, which leads to a large
dispersion of single-age estimates. Conversely, at high values of
$r$ we find 
systems of equal masses with high $r$, whose age estimates are more precise
and lead to lower critical values, and
unbalanced systems for which the age errors and the critical values are greater.

From the analysis of the tables in Appendix~\ref{sec:appcrit}, it appears that
the effect 
of the initial metallicity [Fe/H] is modest, with a mean variation of $W_{0.95}$
by about 0.03 for a change of 1.0 dex in [Fe/H]. This explains the choice 
of grouping the binary systems according to their initial metallicity, neglecting  the change in chemical composition owing to the microscopic diffusion. 

Some words of caution are needed. First of all, the computed critical values directly depend on the assumed magnitude of the
observational uncertainties. 
A larger uncertainty produces a stronger fluctuation in age estimates and thus critical values higher than those presented here. Therefore we calibrated 
the uncertainties we adopted here by assuming realistic values, which are slightly higher than the average of the quoted uncertainties 
in some recent determinations \citep[e.g.][]{Yildiz2007,Clausen2009,Clausen2010,Southworth2013,Torres2014}. 

Moreover, we provide an on-line tool\footnote{\url{http://astro.df.unipi.it/stellar-models/W/}.} that allows computing the required critical value for the supplied masses, 
metallicity, evolutionary phase, and observational uncertainties. This calculator can be useful when the uncertainties on the binary system observables are larger than those adopted here. 

Another aspect that is worth discussing is whether the critical values strongly depend on the adopted stellar models. If there is no dependence, 
the critical values we computed here can be readily used regardless of the stellar models used for the age estimation. If the values
do depend on the models, these critical values 
can be safely adopted only when stellar ages are determined by means of the SCEPtER grid. To answer this question, it is not possible to simply use the stellar model grids 
currently available in literature because they are too sparse. The only way is to compute fine grids of models covering the same parameter space 
as that of SCEPtER by means of different evolutionary codes. Unfortunately, only one code is freely available: the code MESA \citep{MESA2013}. 
We therefore computed a grid of stellar models assuming default MESA input with the following exceptions: solar heavy-element mixtures from \citet{AGSS09}; 
solar-calibrated mixing-length $\alpha_{\rm ml}$ = 1.8; including the element diffusion with the coefficients by \citet{thoul94} with radiation turbulence by \citet{Morel2002}; and the $^{14}$N$(p,\gamma)^{15}$O rate from \citet{Imbriani2004}. 
Then we repeated the previously described steps (see Fig.~\ref{fig:schema1}) to compute the MESA-based critical values. 

The comparison between FRANEC- and MESA-based critical values is quite encouraging because it shows only small variations. The median 
of the differences between FRANEC and MESA was 0.009, with 16th to 84th quantile range [$-0.002$; 0.032]. 
Although a more systematic exploration with other widely used stellar evolution codes would be worthwhile, this first comparison 
suggests that the critical values provided by our procedure are generally applicable. 

SCEPtER age estimates for single stars and binary systems can
be easily obtained  from the R libraries {\it SCEPtER}\footnote{\url{http://CRAN.R-project.org/package=SCEPtER}} and {\it SCEPtERbinary}\footnote{\url{http://CRAN.R-project.org/package=SCEPtERbinary}} that are accessible on CRAN.

\section{Comparison of the $W$ test with the individual age confidence intervals}\label{sec:Wint}

The classical way to asses the stellar coevality in a binary system is to 
fit the observations with isochrones of different ages and claim coevality whenever a single isochrone is able to fit -- within the errors -- the whole system. 
Equivalently, it is possible to estimate the two stellar ages and their errors and establish the coevality by the overlap of the two error intervals. 

A natural question is therefore how the $W$ test performs with respect to the more intuitive approach of adopting the age error intervals to claim system coevality. 
The purpose of this section is to verify whether the two techniques are equivalent. In other words, does the rejection of the coevality 
by the $W$ test imply that the ages of the single members are not compatible with each other within the errors and vice versa?

\begin{figure*}
        \centering
        \includegraphics[width=17.5cm,angle=0]{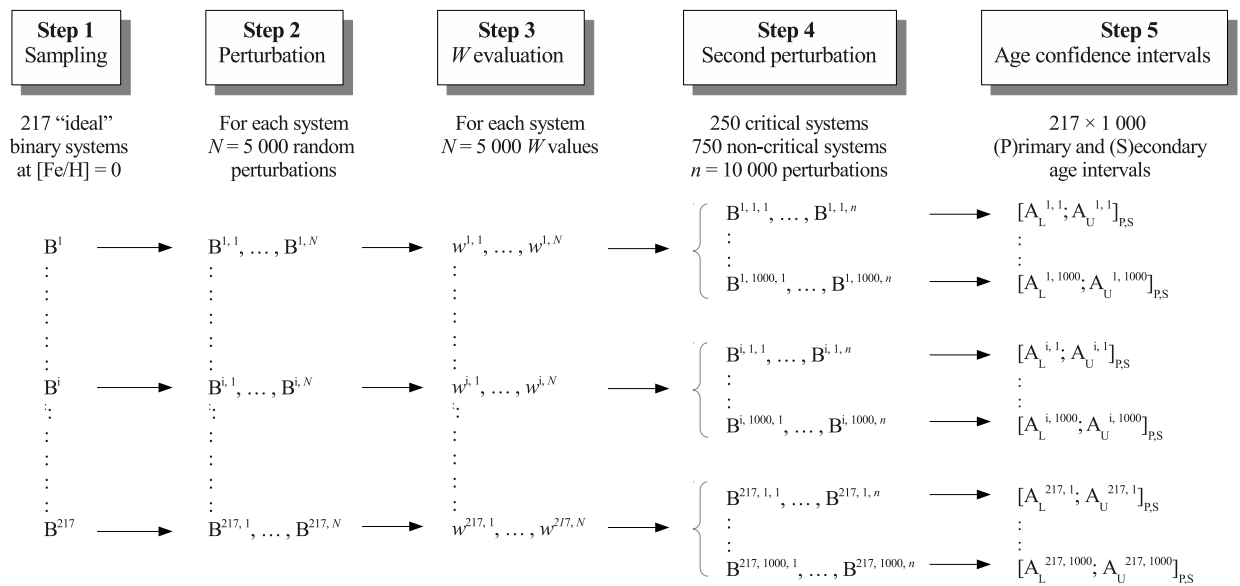}
        \caption{Outline of the comparison between the coevality tests based on $W$ and confidence intervals (see text).}
        \label{fig:schema2}
\end{figure*}

The outline of the procedure is presented in Fig.~\ref{fig:schema2}. It required a huge amount of Monte Carlo simulations, lasting for 12 days on a Intel Xenon machine with 32 cores. To reduce the computational burden, we restricted the calculations to solar metallicity; 217 binary systems from the possible 1087 of Sect.~\ref{sec:sampling} were entered in the analysis. 

As a first step, we generated $N = 5\,000$ Monte Carlo perturbed systems for each of the 217 possible couples (step 2 in Fig.~\ref{fig:schema2})
for the selected 217 ideal binary systems. We estimated the ages of the two stars for each of these $N$ systems and computed the $W$ values (step 3). For the 5\% of these perturbed systems whose values were greater than the critical $W$ (i.e. 250 systems for each of the 217 ideal binaries) we computed the Monte Carlo 95\% confidence interval for the age. That is, for each of these critical systems we generated $n = 10\,000$ newly perturbed systems, estimated the two stellar ages, and obtained the 95\% confidence interval on the ages by computing the 2.5th and the 97.5th quantiles of the age estimates \citep[see][for details]{eta}. The same procedure was repeated for a random set (about 15\% of the total, for computational reasons) of systems with $W$ lower than the critical values (steps 4 and 5 in Fig.~\ref{fig:schema2}).
The procedure required a total of $5.4 \times 10^8$ binary age estimates.
 
The results of these computations showed only a few systems for which the $W$ was lower than the critical values, but the age confidence intervals did not overlap. We can therefore safely disregard these cases.  
This means that whenever the coevality of the binary components cannot be rejected by the $W$ test, the individual stellar ages are compatible with each other within the errors. 

\begin{table*}[ht]
        \centering
        \caption{Medians and ranges from the 16th to 84th quantiles (in parentheses) of the percentages of non-overlapping 95\% confidence intervals for age estimates for systems with $W$ greater than the corresponding critical value, classified according to the relative age of the primary star and the mass ratio of the system.}
        \label{tab:W-intconf}
        \begin{tabular}{ccccc}
                \hline\hline
                \multicolumn{5}{c}{relative age}\\      
                0.1 & 0.3 & 0.5 & 0.7 & 0.9\\
                \hline
                0.0 & 1.6 & 1.2 & 1.2 & 0.4\\
                (0.0; 2.8) &  (0.0; 5.2) &  (0.0; 9.2) &  (0.0; 14.8) & (0.0; 10.4)\\
                \hline
                \multicolumn{5}{c}{mass ratio}\\        
                $[0.5; 0.6]$ & $[0.6; 0.7]$ & $[0.7; 0.8]$ & $[0.8; 0.9]$ & $[0.9; 1.0]$\\
                \hline
                16.2   &   11.0   &   4.4   &   0.4  &     0.0\\
                (9.5;19.6) & (9.1; 18.8) & (2.0; 7.6) & (0.0; 1.2) & (0.0; 0.0)\\
                \hline
        \end{tabular}
\end{table*}

Conversely, we found a striking lack of agreement between the two techniques for systems with $W$ greater than the critical values. Only in a small fraction of cases the comparison of the age confidence intervals revealed a significant difference in the two stellar ages. 
In the vast majority of cases in which the W test allows rejecting coevality, the estimates of the individual ages are still compatible with each other within the errors. 
Under the null hypothesis $H_0$, the confidence interval approach is therefore more conservative for the coevality assessment than the $W$ test, and adopting it results in a severe underestimation of the fraction of systems for which the coevality hypothesis is questionable.  
In summary, the $W$ test and the confidence interval computation have a different scope of applicability. Whenever the former is specifically developed for the task of a coevality check, the confidence interval computation lack of statistical power when adopted for this aim and its actual level is about an order of magnitude lower than the nominal $\alpha = 0.05$.
 
The results for the 217 considered systems are summarised in Table~\ref{tab:W-intconf}, which contains the median and the 16th and 84th quantiles of the percentage of systems for which the $W$ test and the confidence interval techniques concordantly report a significant difference between the ages of the two stars.      

We found decreasing trends with increasing mass ratio of the system; while for systems with $0.5 \leq q \leq 0.6$ the confidence interval method rejects the coevality hypothesis for a median fraction of 16.2\%  of the systems that are significant according
to the $W$ test, this percentage drops to zero for systems with near equal masses. This trend is expected and is mainly due to the assumed correlations in the perturbation step. Systems with $q \approx 1$ , which are very near in the grid before the perturbation, will still be near after the perturbation. Therefore their age confidence intervals will always overlap.   

\section{Toward the $W$ test application on real binary systems}

Another problem deserving a detailed analysis before adopting the $W$ test for real binary systems concerns measurement errors. The empirically determined values $\hat{\cal{S}} = \{ \hat{T}_{\rm eff}^{1,2}, \hat{{\rm [Fe/H]}^{1,2}}, \hat{M}^{1,2}, \hat{R}^{1,2}\}$ of the physical quantities of the two stellar components 
used in the age estimate procedure are in principle different from the real ones ${\cal{S}} = \{ T_{\rm eff}^{1,2}, {\rm [Fe/H]^{1,2}}, M^{1,2}, R^{1,2} \}$. 
Moreover, the evolutionary stage $r$ of the primary star is not observable and has to be estimated. 
On the other hand, we showed in Sect.~\ref{sec:results} that the critical values ${W}_{0.95}$ depends on the true values of stellar masses, metallicity, and primary relative age $ r $. 
As a consequence, the critical value $\hat{W}_{0.95}$ computed following the five steps described in Sect.~\ref{sec:method} starting from the empirically 
determined values $\hat{\cal{S}} $ of the binary members is in principle different from the true critical value ${W}_{0.95}$ computed from the real $\cal{S}$ values. 
Because of this situation, the recovered $W$ value for the observed system should not 
be compared with $\hat{W}_{0.95}$ but with ${W}_{0.95}$. The true $\cal{S}$ values are not observables, however, and thus ${W}_{0.95}$ 
cannot be directly computed. Which values of masses, metallicity, and evolutionary stage $ r $ should be adopted to properly 
compute a critical value that is a satisfactory approximation of the true ${W}_{0.95}$? 
This is a crucial question; the $W$ test can only be applied to real systems if the correct critical value to be adopted in the age comparison can be accurately recovered.
   
The problem is equivalent to finding the most probable true binary system associated with the observed one, or in other words, a system 
composed of two coeval members that maximizes the likelihood of generating the observed binaries after a perturbation caused
by the 
observational uncertainties. \citet[][]{binary} showed that the best solution to this problem is provided by 
imposing the coevality of the two stars in the grid-based recovery procedure. Let $\tilde{\cal{S}}= \{ \tilde{T_{\rm eff}}^{1,2}, \tilde{{\rm [Fe/H]}^{1,2}}, \tilde{M}^{1,2}, \tilde{R}^{1,2} \}$ 
be the best estimate of $\cal{S}$ under this assumption. We therefore performed a new set of Monte Carlo simulations to compute the 
critical values $\tilde{{W}_{0.95}}$ from $\tilde{\cal{S}}$. To check the goodness of this approach, we tested it on a sample  
of synthetic binary systems for which the true values $\cal{S}$ and ${W}_{0.95}$ are known. Then the comparison
between ${W}_{0.95}$ and $\tilde{{W}}_{0.95}$ will prove the performance of the adopted procedure.  

Adopting the same framework as described above,  we obtained the best estimates $\tilde{\cal{S}}$ and 
the corresponding critical values $\tilde{{W}}_{0.95}$ for all the systems
relevant to the $W$ test (critical systems in the step 4 of
Fig.~\ref{fig:schema2}). A possible complication arises whenever the two stars have age estimates so different that no grid-based coeval solutions exist, respecting the observational constraints. 
For these extreme cases the single-star estimates can be adopted. It is clear, however, that the impossibility of obtaining a grid-based coeval solution strongly advises against this hypothesis.     

As a result, we found that the the proposed estimator of the true critical value computed starting from the most probable coeval 
binary system associated with the observed one is good, that
is, unbiased and with a small variance. The differences between $\tilde{W}_{0.95}$ and ${W}_{0.95}$ are small because overall the median 
difference is $-0.004$ (16th and 84th quantile, $-0.019$ and $0.018, $ respectively). No relevant trends with the mass of the stars, the evolutionary phase, or the mass ratio were found. To quantify these variations in terms of the level of the test, they correspond to a median $\alpha$ of 0.048, (16th and 84th quantiles, 0.033 and 0.066).

In conclusion, the Monte Carlo simulations showed that the test can be safely applied to real systems and is more sensitive in rejecting the coevality than the simple computation of the individual age confidence intervals. To adopt the test in practice, it is
therefore necessary to \begin{enumerate}
\item compute the two single-age estimates and the $W$ value; 
\item obtain the best coeval solution of the system; 
\item adopt the masses, metallicity, and relative age of the primary stars obtained in the preceding step to find the critical value $W_{0.95}$ by interpolating the Tables~\ref{tab:crit0.1} to  \ref{tab:crit0.9}; and 
\item  compare the $W$ value with $W_{0.95}$. 
\end{enumerate}

\section{Analytical approximation of the $W$ distribution}\label{sec:distrib}  

For all the combinations of masses, metallicities, and relative ages described in
Sect.~\ref{sec:method}, the 
Gaussian perturbations -- added to the values of the observables before the age
estimation -- cause a scatter of the values of $W$. The actual distribution of
these values ultimately depends on the position in the estimation grid of the
unperturbed values. Since the grid is irregular and the differences among near
models change with the models evolutionary stage \citep[see][for a detailed
  discussion]{scepter1}, it is impossible to explicitly derive  the exact
distributions of $W$ for all the examined cases.

Nevertheless, we were able to find an empirical approximation suitable for all these distributions. 
In this section we show that these distributions are closely
approximated by beta distributions and discuss the validity and limits
of this 
approximation. This result is particularly useful since it
could be used to estimate the critical
values at different levels $\alpha$.

The beta density function $f(x,a, b)$ is defined on the interval $x \in [0, 1]$ and is parametrised by 
two positive parameters, $a$ and $b$, controlling the shape of the
distribution. The density has the expression
\begin{equation}
f(x,a, b) = \frac{\Gamma(a + b)}{\Gamma(a)
  \Gamma(b} x^{a -1} \left(1-x\right)^{b-1}, 
\label{eq:beta}
\end{equation}
where $\Gamma(.)$ is the gamma function. The mean $\mu$ and the variance
$\sigma^2$ of the
distribution are
\begin{eqnarray}
\mu & = & \frac{a}{a+b} \nonumber\\
\sigma^2 & = &\frac{a \; b}{\left(a + b\right)^2 \left(a +
  b+1\right)}.
\label{eq:batams}
\end{eqnarray}

\begin{figure*}
\centering
\includegraphics[height=17cm,angle=-90]{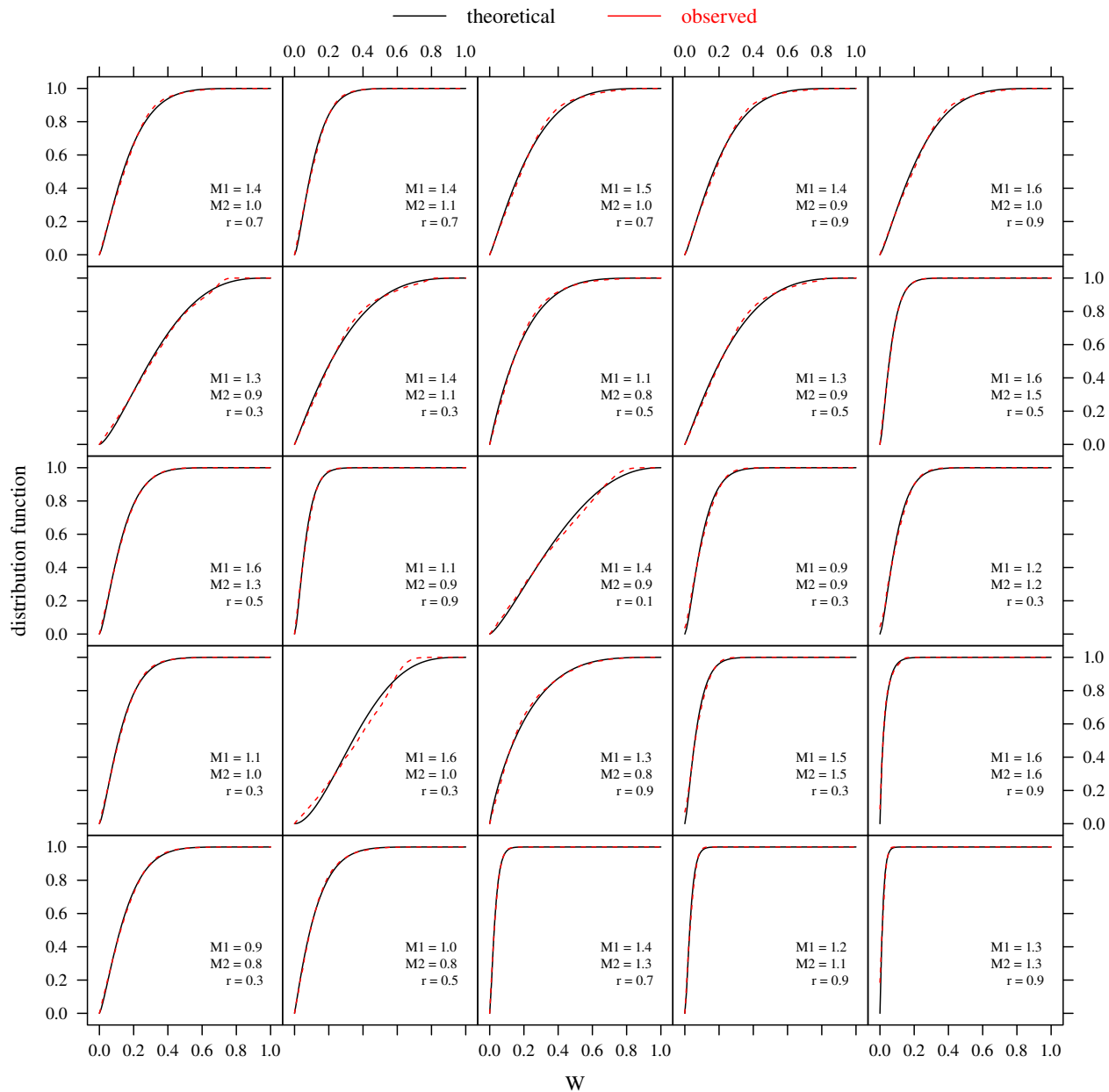}
\caption{Comparison of beta distribution functions (solid black line) with the
  observed distributions (red dashed line) of $W$ for 25 randomly selected
  combinations of masses $M_1$ and $M_2$ of stars (in solar units) and relative age $r$ of the
  primary star (see text).}
\label{fig:beta}
\end{figure*}

For each of the 1087 simulated binary systems of Sect.~\ref{sec:results} we adapted a
beta distribution to the 
$N = 50\,000$ synthetic values of $W$. The parameters of the distribution were
computed by equating the sample mean and variance to the theoretical values
given by Eq.~(\ref{eq:batams}). Figure~\ref{fig:beta} shows a comparison
between the theoretical and the sample distributions for 25 randomly selected
binary systems. The agreement is surprisingly good. 
All the computed critical values, their errors, and the parameters of the beta
approximating distribution function are available in electronic format at
CDS. Table~\ref{tab:totres} shows the first four lines of the table.

\begin{figure}
\centering
\includegraphics[height=8cm,angle=-90]{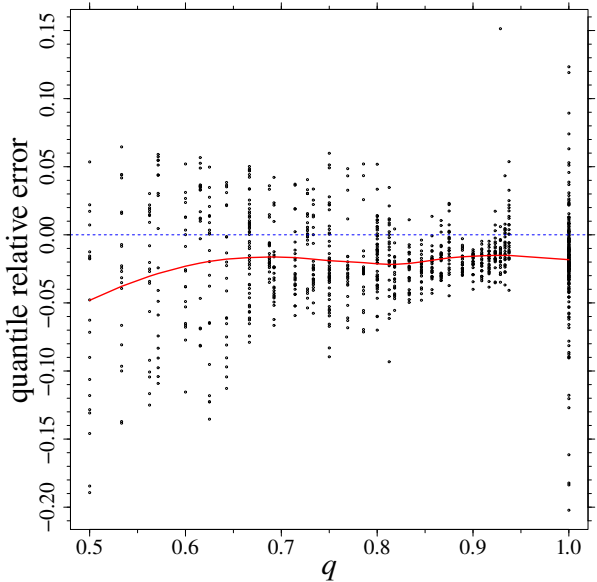}
\caption{Relative difference of the $W_{0.95}$ critical values obtained from the sample
data and from the approximating beta distribution, in dependence on the mass
ratio $q$ (see text). A positive value
implies that the computed quantile is larger than the theoretical one. The red
line is a loess smoother of the data that shows the mean trend of the relative
error.}
\label{fig:quantbeta}
\end{figure}

In Fig.~\ref{fig:quantbeta} we show, in dependence on the mass ratio $q$, the
value $(W_{0.95}-B_{0.95})/W_{0.95}$,  where $W_{0.95}$ are the values computed in Sect.~\ref{sec:results}, and $B_{0.95}$ are the corresponding quantiles obtained from the theoretical beta
distribution. A
positive value means that the sample quantile is larger than the
corresponding theoretical value. To better show the median trend of the
relative error, the figure also shows a loess smoother\footnote{A loess 
regression smoother is 
a non-parametric  locally weighted polynomial regression technique that is
often used to show the underlying trend of scattered data \citep[see
e.g.][]{Feigelson2012,venables2002modern}.} of the data. The data and theoretical estimates agree very well for $q >
0.7$, where the spread is lower than about $\pm5\%$. For lower $q$ there are
larger 
differences and the beta distribution overestimates the desired
quantile by as much as 15\%. A global weak bias is present and the theoretical
quantiles generally overestimate the observed ones, providing a more
conservative test. The median 
differences among theoretical and 
empirical quantiles, marked by the loess smoother, are of about $-2\%$ at high
$q$, and reach a $-5\%$ at $q = 0.5$.

\begin{figure}
\centering
\includegraphics[height=8cm,angle=-90]{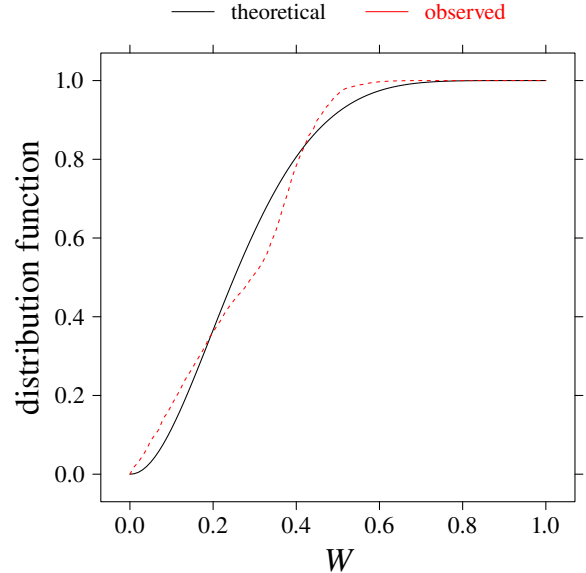}
\caption{Comparison of beta distribution function (solid black) with the
  observed distribution (dashed red) for the case of worst agreement ($M_1$ =
  1.6 
  $M_{\sun}$, $M_2$ = 0.8 $M_{\sun}$, $r$ = 0.5).}
\label{fig:beta-bad}
\end{figure}

Figure~\ref{fig:quantbeta} shows that the variability in the differences between the observed and theoretical quantiles is higher at $q = 0.5$. The typical behaviour for these systems is shown in Fig.~\ref{fig:beta-bad}, which displays the case of worst agreement between
theoretical and empirical distributions ($M_1$ = 1.6
  $M_{\sun}$, $M_2$ = 0.8 $M_{\sun}$, $r$ = 0.5). The empirical distribution
shows a lack of values around $W = 0.3$ and an accumulation at $W = 0.5$.

\begin{table*}[ht]
\centering
\caption{First four lines of the table available at CDS, containing the
  $W_{0.95}$ critical values and the parameters of the beta distribution, which approximate the $W$ distribution,  for the 1087 considered binary systems.}
\label{tab:totres}
\begin{tabular}{cccccccc}
  \hline\hline
 $M_1$ & $M_2$ & [Fe/H] & $r$ & $W_{0.95}$ & $\sigma(W_{0.95})$ & $a$
  & $b$\\
  \hline
  0.8 & 0.8 &-0.55 &  0.1 & 0.5911 & 0.0016 &1.120367 &3.617874\\
  0.9 & 0.8 &-0.55 &  0.1 & 0.6430 & 0.0022 &1.502503 &3.389753\\
  0.9 & 0.9 &-0.55 &  0.1 & 0.5607 & 0.0026 &1.225152 &3.921876\\
  1.0 & 0.8 &-0.55 &  0.1 & 0.6255 & 0.0028 &1.529220 &3.580155\\
\multicolumn{8}{c}{\ldots}\\
\hline
\end{tabular}
\tablefoot{The columns report the mass of the primary star, the mass of the
  secondary, the initial metallicity [Fe/H] of the system, the relative age
  $r$ of the primary star, the critical value $W_{0.95}$, its bootstrap error (see text),
and the parameters $a$ and $b$ of the approximating beta distribution. }
\end{table*}

\section{Conclusions}\label{sec:conclusions}

We devised a statistical test on the difference in the estimated ages of two 
coeval stars in a binary system as a result of the fluctuations caused by
 observational errors. This test allows assessing on statistical grounds 
 whether the apparent non-coevality of binary members is merely 
 the consequence of observational uncertainties. We also provided an on-line tool 
 to be used for real systems. 

We introduced the statistics $W$, defined as the absolute value of the
difference between the two estimated ages and the age of the older star. 
We studied how the $W$ values are scattered
as a result of the uncertainty on the observational constraints adopted in the age
estimation procedure.  
We assumed a level of the statistical test $\alpha = 0.05$, corresponding to critical values $W_{0.95}$. 
The coevality hypothesis is rejected when $W > W_{0.95}$. We analysed the dependence of $W_{0.95}$ on 
the masses of the two stars, the initial metallicity [Fe/H], and the relative age of the primary star. 

We found that the values $W_{0.95}$ range in median from 0.65 for relative age
$r = 0.1$ to 0.2 at relative age $r = 0.9$, meaning that the younger the system, the larger 
the expected difference between the estimated ages of the two components that is due to the observational uncertainties. 
Moreover, $W_{0.95}$, and thus the expected age discrepancy, also increases with the mass of the primary
star. The dependence on the initial metallicity is negligible. 

We also verified that the results are robust to a change of the adopted stellar evolution code. 
To this purpose,  we repeated the process by using a grid of stellar models computed by the MESA evolutionary code \citep{MESA2013}. The median difference in the critical values was about 0.01.

The magnitude of the critical values, in particular for systems near the ZAMS, should
be taken into account in the analysis of observational data before concluding that the
coevality of the stars cannot be accounted for by standard stellar models without changes in the input
physics and/or in the adopted calibration of the free parameters, such as the mixing-length or the convective core overshooting. 

We demonstrated how the test can be adapted to real binary systems in presence of measurements errors on the observed quantities. We showed that the true critical values, computed with perfect knowledge of the observables, are closely approximated by the corresponding values derived by assuming the grid-based best estimates of the masses   imposing the coevality of the solution.
We compared the performance of the $W$ test in assessing the (non-)coevality of the binary components 
with the common approach, which relies on confidence intervals of the individual age estimates. 
We found that the latter approach is too conservative for the assumed level $\alpha$, meaning that most of the systems 
signalled to be non-coeval by the $W$ test have estimated individual ages that are compatible with each other within the errors. In
contrast to the $W$ test, which is specifically developed for the task of a coevality check, the confidence interval comparison is not statistically powerful when adopted for this aim, and it has an actual level of about an order of magnitude lower than the nominal $\alpha = 0.05$.
More in detail, the common approach gives significant differences ranging from about 16\% of the systems with a significant $W$ value at $q \approx 0.5$ to 0\% for systems at $q \approx 1$. 

Finally, we showed that the distributions of $W$ for the various combinations of
star masses, metallicities, and primary relative ages are approximated by beta
distributions with appropriate shape parameters. The approximation is very
good for systems with a mass ratio higher than 0.7, while it is less accurate for
more unbalanced systems.

\begin{acknowledgements}
We thank our referee for the useful comments that helped us in clarifying and improving this paper.
This work has been supported by PRIN-MIUR 2010-2011 ({\em Chemical and dynamical evolution 
        of the Milky Way and Local Group galaxies}, PI F. Matteucci), and  PRIN-INAF 2012 
({\em The M4 Core Project with Hubble Space Telescope}, PI
L. Bedin). 
\end{acknowledgements}

\bibliographystyle{aa}
\bibliography{biblio}

\appendix

\section{Tables of critical values}\label{sec:appcrit}
\begin{table*}[ht]
\centering
\caption{Critical values $W_{0.95}$ in dependence on masses of the stars $M_1$
and $M_2$ (both in solar units), and on their initial metallicity [Fe/H]. Values are computed for primary relative age $r = 0.1$.}
\label{tab:crit0.1}
\begin{tabular}{cc|rrrrrrrrr}
  \hline\hline
 $M_1$ & [Fe/H] & \multicolumn{9}{c}{$M_2$}\\
   &  & 0.8 & 0.9 & 1.0 & 1.1 & 1.2 & 1.3 & 1.4 & 1.5 & 1.6 \\ 
  \hline
   0.80 & -0.55 & 0.591 &  &  &  &  &  &  &  &  \\ 
   0.90 & -0.55 & 0.643 & 0.561 &  &  &  &  &  &  &  \\ 
   1.00 & -0.55 & 0.625 & 0.595 & 0.517 &  &  &  &  &  &  \\ 
   1.10 & -0.55 & 0.617 & 0.603 & 0.581 & 0.536 &  &  &  &  &  \\ 
   1.20 & -0.55 & 0.683 & 0.692 & 0.641 & 0.620 & 0.545 &  &  &  &  \\ 
   1.30 & -0.55 &  & 0.786 & 0.741 & 0.667 & 0.611 & 0.588 &  &  &  \\ 
   1.40 & -0.55 &  & 0.816 & 0.763 & 0.663 & 0.617 & 0.594 & 0.579 &  &  \\ 
   1.50 & -0.55 &  &  & 0.792 & 0.695 & 0.601 & 0.594 & 0.587 & 0.568 &  \\ 
   1.60 & -0.55 &  &  &  & 0.692 & 0.564 & 0.554 & 0.582 & 0.605 & 0.472 \\ 
\hline
   0.80 & -0.25 & 0.598 &  &  &  &  &  &  &  &  \\ 
   0.90 & -0.25 & 0.662 & 0.585 &  &  &  &  &  &  &  \\ 
   1.00 & -0.25 & 0.661 & 0.629 & 0.572 &  &  &  &  &  &  \\ 
   1.10 & -0.25 & 0.629 & 0.606 & 0.603 & 0.534 &  &  &  &  &  \\ 
   1.20 & -0.25 & 0.623 & 0.619 & 0.613 & 0.603 & 0.507 &  &  &  &  \\ 
   1.30 & -0.25 &  & 0.726 & 0.699 & 0.688 & 0.614 & 0.539 &  &  &  \\ 
   1.40 & -0.25 &  &  & 0.779 & 0.734 & 0.682 & 0.593 & 0.574 &  &  \\ 
   1.50 & -0.25 &  &  & 0.796 & 0.740 & 0.676 & 0.592 & 0.574 & 0.562 &  \\ 
   1.60 & -0.25 &  &  &  & 0.732 & 0.650 & 0.582 & 0.576 & 0.601 & 0.458 \\ 
\hline
   0.80 & 0.00 & 0.602 &  &  &  &  &  &  &  &  \\ 
   0.90 & 0.00 & 0.659 & 0.603 &  &  &  &  &  &  &  \\ 
   1.00 & 0.00 & 0.673 & 0.645 & 0.598 &  &  &  &  &  &  \\ 
   1.10 & 0.00 & 0.681 & 0.637 & 0.628 & 0.588 &  &  &  &  &  \\ 
   1.20 & 0.00 & 0.702 & 0.648 & 0.615 & 0.610 & 0.541 &  &  &  &  \\ 
   1.30 & 0.00 &  & 0.686 & 0.623 & 0.628 & 0.604 & 0.518 &  &  &  \\ 
   1.40 & 0.00 &  &  & 0.730 & 0.696 & 0.674 & 0.618 & 0.535 &  &  \\ 
   1.50 & 0.00 &  &  & 0.806 & 0.767 & 0.710 & 0.677 & 0.571 & 0.544 &  \\ 
   1.60 & 0.00 &  &  &  & 0.774 & 0.688 & 0.670 & 0.592 & 0.594 & 0.455 \\ 
\hline
   0.80 & 0.25 & 0.625 &  &  &  &  &  &  &  &  \\ 
   0.90 & 0.25 & 0.683 & 0.618 &  &  &  &  &  &  &  \\ 
   1.00 & 0.25 & 0.678 & 0.660 & 0.620 &  &  &  &  &  &  \\ 
   1.10 & 0.25 & 0.678 & 0.654 & 0.648 & 0.608 &  &  &  &  &  \\ 
   1.20 & 0.25 & 0.708 & 0.658 & 0.630 & 0.633 & 0.584 &  &  &  &  \\ 
   1.30 & 0.25 &  & 0.717 & 0.644 & 0.632 & 0.606 & 0.535 &  &  &  \\ 
   1.40 & 0.25 &  &  & 0.688 & 0.631 & 0.616 & 0.577 & 0.534 &  &  \\ 
   1.50 & 0.25 &  &  & 0.769 & 0.713 & 0.674 & 0.632 & 0.595 & 0.510 &  \\ 
   1.60 & 0.25 &  &  &  & 0.780 & 0.729 & 0.677 & 0.661 & 0.586 & 0.445 \\ 
\hline
   0.80 & 0.55 & 0.643 &  &  &  &  &  &  &  &  \\ 
   0.90 & 0.55 & 0.693 & 0.644 &  &  &  &  &  &  &  \\ 
   1.00 & 0.55 & 0.687 & 0.671 & 0.630 &  &  &  &  &  &  \\ 
   1.10 & 0.55 & 0.671 & 0.653 & 0.659 & 0.620 &  &  &  &  &  \\ 
   1.20 & 0.55 & 0.715 & 0.653 & 0.649 & 0.647 & 0.604 &  &  &  &  \\ 
   1.30 & 0.55 &  & 0.699 & 0.643 & 0.633 & 0.630 & 0.580 &  &  &  \\ 
   1.40 & 0.55 &  & 0.731 & 0.659 & 0.611 & 0.630 & 0.626 & 0.495 &  &  \\ 
   1.50 & 0.55 &  &  & 0.708 & 0.643 & 0.592 & 0.601 & 0.537 & 0.421 &  \\ 
   1.60 & 0.55 &  &  & 0.748 & 0.685 & 0.612 & 0.610 & 0.619 & 0.559 & 0.281 \\ 
   \hline
\end{tabular}
\end{table*}

\begin{table*}[ht]
\centering
\caption{As in Table~\ref{tab:crit0.1}, but for a primary relative age $r = 0.3$.}
\label{tab:crit0.3}
\begin{tabular}{cc|rrrrrrrrr}
  \hline\hline
 $M_1$ & [Fe/H] & \multicolumn{9}{c}{$M_2$}\\
   &  & 0.8 & 0.9 & 1.0 & 1.1 & 1.2 & 1.3 & 1.4 & 1.5 & 1.6 \\ 
  \hline
  0.80 & -0.55 & 0.230 &  &  &  &  &  &  &  &  \\ 
  0.90 & -0.55 & 0.358 & 0.229 &  &  &  &  &  &  &  \\ 
  1.00 & -0.55 & 0.521 & 0.310 & 0.215 &  &  &  &  &  &  \\ 
  1.10 & -0.55 & 0.638 & 0.448 & 0.291 & 0.204 &  &  &  &  &  \\ 
  1.20 & -0.55 & 0.618 & 0.592 & 0.423 & 0.288 & 0.200 &  &  &  &  \\ 
  1.30 & -0.55 & 0.547 & 0.651 & 0.539 & 0.401 & 0.265 & 0.209 &  &  &  \\ 
  1.40 & -0.55 & 0.461 & 0.604 & 0.602 & 0.489 & 0.345 & 0.288 & 0.210 &  &  \\ 
  1.50 & -0.55 & 0.468 & 0.628 & 0.614 & 0.600 & 0.469 & 0.435 & 0.310 & 0.210 &  \\ 
  1.60 & -0.55 & 0.483 & 0.674 & 0.589 & 0.672 & 0.605 & 0.563 & 0.465 & 0.345 & 0.162 \\ 
\hline
  0.80 & -0.25 & 0.231 &  &  &  &  &  &  &  &  \\ 
  0.90 & -0.25 & 0.367 & 0.238 &  &  &  &  &  &  &  \\ 
  1.00 & -0.25 & 0.542 & 0.331 & 0.230 &  &  &  &  &  &  \\ 
  1.10 & -0.25 & 0.662 & 0.487 & 0.318 & 0.217 &  &  &  &  &  \\ 
  1.20 & -0.25 & 0.646 & 0.629 & 0.482 & 0.303 & 0.195 &  &  &  &  \\ 
  1.30 & -0.25 & 0.557 & 0.665 & 0.628 & 0.449 & 0.275 & 0.204 &  &  &  \\ 
  1.40 & -0.25 & 0.484 & 0.632 & 0.676 & 0.571 & 0.370 & 0.258 & 0.209 &  &  \\ 
  1.50 & -0.25 & 0.467 & 0.622 & 0.658 & 0.642 & 0.449 & 0.351 & 0.285 & 0.202 &  \\ 
  1.60 & -0.25 & 0.482 & 0.655 & 0.639 & 0.684 & 0.551 & 0.485 & 0.447 & 0.303 & 0.158 \\ 
\hline
  0.80 & 0.00 & 0.236 &  &  &  &  &  &  &  &  \\ 
  0.90 & 0.00 & 0.358 & 0.244 &  &  &  &  &  &  &  \\ 
  1.00 & 0.00 & 0.527 & 0.340 & 0.241 &  &  &  &  &  &  \\ 
  1.10 & 0.00 & 0.632 & 0.505 & 0.328 & 0.232 &  &  &  &  &  \\ 
  1.20 & 0.00 & 0.606 & 0.628 & 0.487 & 0.319 & 0.217 &  &  &  &  \\ 
  1.30 & 0.00 & 0.551 & 0.662 & 0.638 & 0.480 & 0.298 & 0.196 &  &  &  \\ 
  1.40 & 0.00 & 0.493 & 0.621 & 0.689 & 0.641 & 0.440 & 0.271 & 0.182 &  &  \\ 
  1.50 & 0.00 & 0.514 & 0.633 & 0.661 & 0.705 & 0.560 & 0.379 & 0.226 & 0.184 &  \\ 
  1.60 & 0.00 & 0.537 & 0.667 & 0.639 & 0.717 & 0.664 & 0.473 & 0.348 & 0.265 & 0.149 \\ 
\hline
  0.80 & 0.25 & 0.238 &  &  &  &  &  &  &  &  \\ 
  0.90 & 0.25 & 0.364 & 0.237 &  &  &  &  &  &  &  \\ 
  1.00 & 0.25 & 0.537 & 0.324 & 0.242 &  &  &  &  &  &  \\ 
  1.10 & 0.25 & 0.657 & 0.491 & 0.330 & 0.238 &  &  &  &  &  \\ 
  1.20 & 0.25 & 0.646 & 0.637 & 0.509 & 0.319 & 0.227 &  &  &  &  \\ 
  1.30 & 0.25 & 0.569 & 0.653 & 0.661 & 0.488 & 0.304 & 0.213 &  &  &  \\ 
  1.40 & 0.25 & 0.508 & 0.611 & 0.683 & 0.644 & 0.450 & 0.265 & 0.192 &  &  \\ 
  1.50 & 0.25 & 0.525 & 0.643 & 0.647 & 0.712 & 0.609 & 0.385 & 0.239 & 0.176 &  \\ 
  1.60 & 0.25 &  & 0.701 & 0.639 & 0.716 & 0.694 & 0.526 & 0.356 & 0.237 & 0.147 \\ 
\hline
  0.80 & 0.55 & 0.237 &  &  &  &  &  &  &  &  \\ 
  0.90 & 0.55 & 0.345 & 0.248 &  &  &  &  &  &  &  \\ 
  1.00 & 0.55 & 0.529 & 0.333 & 0.242 &  &  &  &  &  &  \\ 
  1.10 & 0.55 & 0.686 & 0.507 & 0.316 & 0.241 &  &  &  &  &  \\ 
  1.20 & 0.55 & 0.660 & 0.682 & 0.470 & 0.307 & 0.238 &  &  &  &  \\ 
  1.30 & 0.55 & 0.598 & 0.685 & 0.639 & 0.450 & 0.295 & 0.229 &  &  &  \\ 
  1.40 & 0.55 & 0.524 & 0.633 & 0.708 & 0.658 & 0.473 & 0.310 & 0.206 &  &  \\ 
  1.50 & 0.55 & 0.494 & 0.604 & 0.658 & 0.695 & 0.598 & 0.376 & 0.232 & 0.187 &  \\ 
  1.60 & 0.55 & 0.517 & 0.655 & 0.631 & 0.701 & 0.701 & 0.537 & 0.299 & 0.241 & 0.142 \\  
   \hline
\end{tabular}
\end{table*}

\begin{table*}[ht]
\centering
\caption{As in Table~\ref{tab:crit0.1}, but for a primary relative age $r = 0.5$.}
\label{tab:crit0.5}
\begin{tabular}{cc|rrrrrrrrr}
  \hline\hline
 $M_1$ & [Fe/H] & \multicolumn{9}{c}{$M_2$}\\
   &  & 0.8 & 0.9 & 1.0 & 1.1 & 1.2 & 1.3 & 1.4 & 1.5 & 1.6 \\ 
  \hline
 0.80 & -0.55 & 0.129 &  &  &  &  &  &  &  &  \\ 
 0.90 & -0.55 & 0.201 & 0.132 &  &  &  &  &  &  &  \\ 
 1.00 & -0.55 & 0.318 & 0.195 & 0.126 &  &  &  &  &  &  \\ 
 1.10 & -0.55 & 0.474 & 0.288 & 0.180 & 0.117 &  &  &  &  &  \\ 
 1.20 & -0.55 & 0.618 & 0.416 & 0.275 & 0.165 & 0.103 &  &  &  &  \\ 
 1.30 & -0.55 & 0.678 & 0.536 & 0.364 & 0.235 & 0.143 & 0.101 &  &  &  \\ 
 1.40 & -0.55 & 0.608 & 0.619 & 0.455 & 0.315 & 0.193 & 0.143 & 0.113 &  &  \\ 
 1.50 & -0.55 & 0.526 & 0.653 & 0.558 & 0.418 & 0.285 & 0.202 & 0.146 & 0.117 &  \\ 
 1.60 & -0.55 & 0.457 & 0.613 & 0.611 & 0.498 & 0.404 & 0.349 & 0.236 & 0.162 & 0.090 \\ 
\hline
 0.80 & -0.25 & 0.129 &  &  &  &  &  &  &  &  \\ 
 0.90 & -0.25 & 0.208 & 0.138 &  &  &  &  &  &  &  \\ 
 1.00 & -0.25 & 0.337 & 0.196 & 0.132 &  &  &  &  &  &  \\ 
 1.10 & -0.25 & 0.496 & 0.308 & 0.196 & 0.129 &  &  &  &  &  \\ 
 1.20 & -0.25 & 0.647 & 0.455 & 0.299 & 0.183 & 0.112 &  &  &  &  \\ 
 1.30 & -0.25 & 0.698 & 0.599 & 0.427 & 0.274 & 0.156 & 0.098 &  &  &  \\ 
 1.40 & -0.25 & 0.629 & 0.661 & 0.551 & 0.377 & 0.221 & 0.144 & 0.110 &  &  \\ 
 1.50 & -0.25 & 0.541 & 0.676 & 0.652 & 0.459 & 0.280 & 0.192 & 0.148 & 0.113 &  \\ 
 1.60 & -0.25 & 0.477 & 0.643 & 0.690 & 0.553 & 0.366 & 0.288 & 0.222 & 0.158 & 0.092 \\ 
\hline
 0.80 & 0.00 & 0.130 &  &  &  &  &  &  &  &  \\ 
 0.90 & 0.00 & 0.212 & 0.140 &  &  &  &  &  &  &  \\ 
 1.00 & 0.00 & 0.329 & 0.201 & 0.138 &  &  &  &  &  &  \\ 
 1.10 & 0.00 & 0.490 & 0.303 & 0.195 & 0.136 &  &  &  &  &  \\ 
 1.20 & 0.00 & 0.627 & 0.452 & 0.300 & 0.197 & 0.132 &  &  &  &  \\ 
 1.30 & 0.00 & 0.676 & 0.610 & 0.442 & 0.286 & 0.179 & 0.108 &  &  &  \\ 
 1.40 & 0.00 & 0.597 & 0.700 & 0.594 & 0.419 & 0.257 & 0.157 & 0.099 &  &  \\ 
 1.50 & 0.00 & 0.529 & 0.675 & 0.687 & 0.548 & 0.363 & 0.223 & 0.131 & 0.104 &  \\ 
 1.60 & 0.00 & 0.483 & 0.638 & 0.725 & 0.638 & 0.457 & 0.296 & 0.189 & 0.149 & 0.087 \\ 
\hline
 0.80 & 0.25 & 0.132 &  &  &  &  &  &  &  &  \\ 
 0.90 & 0.25 & 0.203 & 0.139 &  &  &  &  &  &  &  \\ 
 1.00 & 0.25 & 0.314 & 0.197 & 0.141 &  &  &  &  &  &  \\ 
 1.10 & 0.25 & 0.499 & 0.289 & 0.194 & 0.139 &  &  &  &  &  \\ 
 1.20 & 0.25 & 0.649 & 0.436 & 0.293 & 0.191 & 0.134 &  &  &  &  \\ 
 1.30 & 0.25 & 0.699 & 0.608 & 0.450 & 0.288 & 0.186 & 0.122 &  &  &  \\ 
 1.40 & 0.25 & 0.630 & 0.701 & 0.615 & 0.418 & 0.256 & 0.161 & 0.109 &  &  \\ 
 1.50 & 0.25 & 0.545 & 0.671 & 0.698 & 0.574 & 0.364 & 0.220 & 0.148 & 0.104 &  \\ 
 1.60 & 0.25 & 0.481 & 0.636 & 0.729 & 0.671 & 0.498 & 0.325 & 0.222 & 0.151 & 0.079 \\ 
\hline
 0.80 & 0.55 & 0.134 &  &  &  &  &  &  &  &  \\ 
 0.90 & 0.55 & 0.210 & 0.144 &  &  &  &  &  &  &  \\ 
 1.00 & 0.55 & 0.308 & 0.202 & 0.141 &  &  &  &  &  &  \\ 
 1.10 & 0.55 & 0.469 & 0.295 & 0.194 & 0.148 &  &  &  &  &  \\ 
 1.20 & 0.55 & 0.663 & 0.441 & 0.278 & 0.191 & 0.147 &  &  &  &  \\ 
 1.30 & 0.55 & 0.709 & 0.636 & 0.406 & 0.273 & 0.187 & 0.136 &  &  &  \\ 
 1.40 & 0.55 & 0.640 & 0.737 & 0.598 & 0.396 & 0.271 & 0.186 & 0.127 &  &  \\ 
 1.50 & 0.55 & 0.567 & 0.696 & 0.696 & 0.542 & 0.348 & 0.240 & 0.156 & 0.112 &  \\ 
 1.60 & 0.55 & 0.494 & 0.652 & 0.748 & 0.639 & 0.459 & 0.323 & 0.201 & 0.167 & 0.082 \\   
   \hline
\end{tabular}
\end{table*}

\begin{table*}[ht]
\centering
\caption{As in Table~\ref{tab:crit0.1}, but for a primary relative age $r = 0.7$.}
\label{tab:crit0.7}
\begin{tabular}{cc|rrrrrrrrr}
  \hline\hline
 $M_1$ & [Fe/H] & \multicolumn{9}{c}{$M_2$}\\
   &  & 0.8 & 0.9 & 1.0 & 1.1 & 1.2 & 1.3 & 1.4 & 1.5 & 1.6 \\ 
  \hline
 1.40 & -0.55 & 0.688 & 0.503 & 0.348 & 0.219 & 0.135 & 0.082 & 0.063 &  &  \\ 
 1.50 & -0.55 & 0.641 & 0.616 & 0.431 & 0.294 & 0.199 & 0.144 & 0.083 & 0.064 &  \\ 
 1.60 & -0.55 & 0.581 & 0.666 & 0.498 & 0.394 & 0.283 & 0.220 & 0.177 & 0.112 & 0.050 \\ 
 0.80 & -0.25 & 0.086 &  &  &  &  &  &  &  &  \\ 
 0.90 & -0.25 & 0.142 & 0.098 &  &  &  &  &  &  &  \\ 
 1.00 & -0.25 & 0.225 & 0.141 & 0.090 &  &  &  &  &  &  \\ 
 1.10 & -0.25 & 0.357 & 0.228 & 0.149 & 0.086 &  &  &  &  &  \\ 
 1.20 & -0.25 & 0.507 & 0.323 & 0.208 & 0.124 & 0.088 &  &  &  &  \\ 
 1.30 & -0.25 & 0.639 & 0.458 & 0.308 & 0.186 & 0.104 & 0.065 &  &  &  \\ 
 1.40 & -0.25 & 0.713 & 0.577 & 0.415 & 0.264 & 0.150 & 0.084 & 0.064 &  &  \\ 
 1.50 & -0.25 & 0.658 & 0.660 & 0.518 & 0.343 & 0.200 & 0.132 & 0.092 & 0.065 &  \\ 
 1.60 & -0.25 & 0.599 & 0.719 & 0.606 & 0.424 & 0.266 & 0.201 & 0.162 & 0.114 & 0.049 \\ 
\hline
 0.80 & 0.00 & 0.083 &  &  &  &  &  &  &  &  \\ 
 0.90 & 0.00 & 0.142 & 0.099 &  &  &  &  &  &  &  \\ 
 1.00 & 0.00 & 0.227 & 0.139 & 0.093 &  &  &  &  &  &  \\ 
 1.10 & 0.00 & 0.351 & 0.219 & 0.137 & 0.087 &  &  &  &  &  \\ 
 1.20 & 0.00 & 0.506 & 0.329 & 0.220 & 0.138 & 0.096 &  &  &  &  \\ 
 1.30 & 0.00 & 0.636 & 0.464 & 0.311 & 0.197 & 0.125 & 0.076 &  &  &  \\ 
 1.40 & 0.00 & 0.692 & 0.606 & 0.440 & 0.291 & 0.177 & 0.100 & 0.070 &  &  \\ 
 1.50 & 0.00 & 0.625 & 0.670 & 0.572 & 0.399 & 0.250 & 0.150 & 0.079 & 0.064 &  \\ 
 1.60 & 0.00 & 0.566 & 0.722 & 0.646 & 0.492 & 0.331 & 0.208 & 0.134 & 0.094 & 0.047 \\ 
\hline
 0.80 & 0.25 & 0.084 &  &  &  &  &  &  &  &  \\ 
 0.90 & 0.25 & 0.139 & 0.093 &  &  &  &  &  &  &  \\ 
 1.00 & 0.25 & 0.217 & 0.138 & 0.092 &  &  &  &  &  &  \\ 
 1.10 & 0.25 & 0.345 & 0.208 & 0.136 & 0.090 &  &  &  &  &  \\ 
 1.20 & 0.25 & 0.526 & 0.305 & 0.213 & 0.135 & 0.087 &  &  &  &  \\ 
 1.30 & 0.25 & 0.660 & 0.440 & 0.301 & 0.199 & 0.127 & 0.086 &  &  &  \\ 
 1.40 & 0.25 & 0.713 & 0.608 & 0.444 & 0.283 & 0.181 & 0.111 & 0.083 &  &  \\ 
 1.50 & 0.25 & 0.654 & 0.678 & 0.595 & 0.397 & 0.250 & 0.155 & 0.093 & 0.076 &  \\ 
 1.60 & 0.25 & 0.595 & 0.718 & 0.671 & 0.509 & 0.346 & 0.233 & 0.155 & 0.102 & 0.057 \\ 
\hline
 0.80 & 0.55 & 0.083 &  &  &  &  &  &  &  &  \\ 
 0.90 & 0.55 & 0.140 & 0.093 &  &  &  &  &  &  &  \\ 
 1.00 & 0.55 & 0.217 & 0.139 & 0.091 &  &  &  &  &  &  \\ 
 1.10 & 0.55 & 0.330 & 0.218 & 0.139 & 0.095 &  &  &  &  &  \\ 
 1.20 & 0.55 & 0.492 & 0.313 & 0.205 & 0.138 & 0.098 &  &  &  &  \\ 
 1.30 & 0.55 & 0.675 & 0.451 & 0.289 & 0.199 & 0.137 & 0.098 &  &  &  \\ 
 1.40 & 0.55 & 0.716 & 0.629 & 0.402 & 0.275 & 0.188 & 0.127 & 0.104 &  &  \\ 
 1.50 & 0.55 & 0.662 & 0.728 & 0.539 & 0.361 & 0.249 & 0.163 & 0.112 & 0.096 &  \\ 
 1.60 & 0.55 & 0.611 & 0.739 & 0.651 & 0.480 & 0.336 & 0.241 & 0.154 & 0.116 & 0.069 \\   
   \hline
\end{tabular}
\end{table*}

\begin{table*}[ht]
\centering
\caption{As in Table~\ref{tab:crit0.1}, but for a primary relative age $r = 0.9$.}
\label{tab:crit0.9}
\begin{tabular}{cc|rrrrrrrrr}
  \hline\hline
 $M_1$ & [Fe/H] & \multicolumn{9}{c}{$M_2$}\\
   &  & 0.8 & 0.9 & 1.0 & 1.1 & 1.2 & 1.3 & 1.4 & 1.5 & 1.6 \\ 
  \hline
 0.80 & -0.55 & 0.057 &  &  &  &  &  &  &  &  \\ 
 0.90 & -0.55 & 0.097 & 0.056 &  &  &  &  &  &  &  \\ 
 1.00 & -0.55 & 0.164 & 0.110 & 0.055 &  &  &  &  &  &  \\ 
 1.10 & -0.55 & 0.254 & 0.166 & 0.094 & 0.064 &  &  &  &  &  \\ 
 1.20 & -0.55 & 0.381 & 0.241 & 0.148 & 0.081 & 0.051 &  &  &  &  \\ 
 1.30 & -0.55 & 0.511 & 0.328 & 0.209 & 0.117 & 0.069 & 0.047 &  &  &  \\ 
 1.40 & -0.55 & 0.620 & 0.418 & 0.279 & 0.170 & 0.108 & 0.115 & 0.039 &  &  \\ 
 1.50 & -0.55 & 0.703 & 0.513 & 0.352 & 0.225 & 0.155 & 0.095 & 0.071 & 0.051 &  \\ 
 1.60 & -0.55 & 0.668 & 0.596 & 0.436 & 0.321 & 0.240 & 0.195 & 0.143 & 0.127 & 0.079 \\ 
\hline
 0.80 & -0.25 & 0.057 &  &  &  &  &  &  &  &  \\ 
 0.90 & -0.25 & 0.105 & 0.062 &  &  &  &  &  &  &  \\ 
 1.00 & -0.25 & 0.171 & 0.111 & 0.060 &  &  &  &  &  &  \\ 
 1.10 & -0.25 & 0.266 & 0.182 & 0.117 & 0.068 &  &  &  &  &  \\ 
 1.20 & -0.25 & 0.403 & 0.253 & 0.161 & 0.101 & 0.066 &  &  &  &  \\ 
 1.30 & -0.25 & 0.539 & 0.365 & 0.241 & 0.144 & 0.086 & 0.045 &  &  &  \\ 
 1.40 & -0.25 & 0.664 & 0.472 & 0.326 & 0.205 & 0.123 & 0.070 & 0.045 &  &  \\ 
 1.50 & -0.25 & 0.721 & 0.595 & 0.425 & 0.280 & 0.182 & 0.139 & 0.132 & 0.113 &  \\ 
 1.60 & -0.25 & 0.695 & 0.669 & 0.515 & 0.362 & 0.240 & 0.207 & 0.179 & 0.162 & 0.093 \\ 
\hline
 0.80 & 0.00 & 0.056 &  &  &  &  &  &  &  &  \\ 
 0.90 & 0.00 & 0.103 & 0.063 &  &  &  &  &  &  &  \\ 
 1.00 & 0.00 & 0.175 & 0.109 & 0.067 &  &  &  &  &  &  \\ 
 1.10 & 0.00 & 0.271 & 0.175 & 0.108 & 0.062 &  &  &  &  &  \\ 
 1.20 & 0.00 & 0.399 & 0.258 & 0.170 & 0.105 & 0.075 &  &  &  &  \\ 
 1.30 & 0.00 & 0.547 & 0.367 & 0.244 & 0.155 & 0.098 & 0.056 &  &  &  \\ 
 1.40 & 0.00 & 0.654 & 0.497 & 0.343 & 0.228 & 0.143 & 0.092 & 0.065 &  &  \\ 
 1.50 & 0.00 & 0.694 & 0.621 & 0.461 & 0.308 & 0.216 & 0.148 & 0.136 & 0.112 &  \\ 
 1.60 & 0.00 & 0.656 & 0.681 & 0.556 & 0.409 & 0.294 & 0.205 & 0.164 & 0.157 & 0.100 \\ 
\hline
 0.80 & 0.25 & 0.056 &  &  &  &  &  &  &  &  \\ 
 0.90 & 0.25 & 0.101 & 0.063 &  &  &  &  &  &  &  \\ 
 1.00 & 0.25 & 0.166 & 0.104 & 0.063 &  &  &  &  &  &  \\ 
 1.10 & 0.25 & 0.263 & 0.167 & 0.104 & 0.064 &  &  &  &  &  \\ 
 1.20 & 0.25 & 0.407 & 0.242 & 0.164 & 0.098 & 0.063 &  &  &  &  \\ 
 1.30 & 0.25 & 0.567 & 0.346 & 0.246 & 0.158 & 0.091 & 0.055 &  &  &  \\ 
 1.40 & 0.25 & 0.668 & 0.468 & 0.335 & 0.227 & 0.154 & 0.119 & 0.090 &  &  \\ 
 1.50 & 0.25 & 0.717 & 0.621 & 0.455 & 0.296 & 0.197 & 0.137 & 0.114 & 0.093 &  \\ 
 1.60 & 0.25 & 0.681 & 0.676 & 0.567 & 0.393 & 0.264 & 0.201 & 0.141 & 0.128 & 0.101 \\ 
\hline
 0.80 & 0.55 & 0.058 &  &  &  &  &  &  &  &  \\ 
 0.90 & 0.55 & 0.103 & 0.060 &  &  &  &  &  &  &  \\ 
 1.00 & 0.55 & 0.163 & 0.104 & 0.062 &  &  &  &  &  &  \\ 
 1.10 & 0.55 & 0.251 & 0.169 & 0.105 & 0.064 &  &  &  &  &  \\ 
 1.20 & 0.55 & 0.376 & 0.248 & 0.158 & 0.101 & 0.065 &  &  &  &  \\ 
 1.30 & 0.55 & 0.535 & 0.353 & 0.238 & 0.160 & 0.104 & 0.058 &  &  &  \\ 
 1.40 & 0.55 & 0.685 & 0.491 & 0.319 & 0.222 & 0.159 & 0.116 & 0.081 &  &  \\ 
 1.50 & 0.55 & 0.723 & 0.633 & 0.411 & 0.286 & 0.201 & 0.137 & 0.105 & 0.080 &  \\ 
 1.60 & 0.55 & 0.688 & 0.700 & 0.516 & 0.370 & 0.272 & 0.191 & 0.143 & 0.119 & 0.060 \\   
   \hline
\end{tabular}
\end{table*}

\end{document}